\documentclass[conference]{IEEEtran}
\usepackage{cite}
\usepackage{algorithmic}
\usepackage{textcomp}
\usepackage[hyphens]{url}
\usepackage{hyperref}

\usepackage{graphicx}% Include figure files
\usepackage{tikz}
\usetikzlibrary{shapes,arrows.meta,arrows,fit}
\usepackage{dcolumn}% Align table columns on decimal point
\usepackage{bm}% bold math
\usepackage{amsmath,amsthm,amsfonts}
\usepackage{bbm}
\usepackage{mathtools}
\usepackage{enumerate}
\usepackage[shortlabels]{enumitem}
\usepackage{xcolor}
\usepackage{textcomp}
\usepackage{gensymb} 
\usepackage{float}
\usepackage{soul}
\usepackage{booktabs}
\usepackage{microtype}
\usepackage{graphicx}
\usepackage{xspace}

\newcommand{\ket}[1]{|#1\rangle}

\newcommand{\ionq}{IonQ}

\newtheorem{theorem}{Theorem}[section]
\newtheorem{corollary}[theorem]{Corollary}
\newtheorem{lemma}[theorem]{Lemma}
\newtheorem{example}[theorem]{Example}

\def\BibTeX{{\rm B\kern-.05em{\sc i\kern-.025em b}\kern-.08em
    T\kern-.1667em\lower.7ex\hbox{E}\kern-.125emX}}

\title{Detecting Qubit-coupling Faults \\
in Ion-trap Quantum Computers} 

\author{
\IEEEauthorblockN{ 
Andrii Maksymov,
Jason Nguyen,
Vandiver Chaplin,
Yunseong Nam,
Igor L. Markov 
\\
\{maksymov,nguenj,chaplin,nam,markov\}.ionq.co
}\\
\IEEEauthorblockA{ %\vspace{2em} 
\large IonQ
}}

\begin{document}
\maketitle
\thispagestyle{plain}
\pagestyle{plain}

%%%%%% -- PAPER CONTENT STARTS-- %%%%%%%%

\begin{abstract}
Ion-trap quantum computers offer a large number of possible qubit couplings, each of which requires individual calibration and can be misconfigured. 
To enhance the duty cycle of an ion trap, we develop a strategy that diagnoses individual miscalibrated couplings using only log-many tests. This strategy is validated on a commercial ion-trap quantum computer, where we illustrate the process of debugging faulty quantum gates. Our methodology provides a scalable pathway towards fault detections on a larger scale ion-trap quantum computers, confirmed by simulations up to 32 qubits. 
\end{abstract}

\section{Introduction}

  Quantum computation~\cite{NC} promises to extend the capabilities of conventional computers in science applications~\cite{Femoco,Heisenberg,LGT}, for hard optimization problems~\cite{VQE,QAOA}, and for sensitive cryptography tasks~\cite{Shor,ECC}. Small and intermediate-scale quantum computers based on supercold superconductors~\cite{IBM,Rigetti} and on moderately cooled trapped ions~\cite{Wright_2019,Honeywell} have now been offered commercially via cloud access for five years, as HW accelerators. Improving their operational uptime requires system stability, effective maintenance operations, and fast diagnosis of common problems.

  Compared to conventional VLSI circuit test, several essential differences can be seen in testing and debugging of quantum circuits. In particular, testing and calibration are performed frequently between production runs as part of routine maintenance. The following executive summary is applicable to most qubit technologies except for photonic qubits.

\begin{itemize}
\item Each qubit represents a unit of storage, and gates are applied to qubits with sequential semantics.
\item Qubits are short-lived and must be re-initialized often.
\item The initial state of each qubit is zero; while an arbitrary bitstring can be configured via inverters, creating an arbitrary quantum state for test purposes is impractical.
\item Quantum states are observed via stochastic destructive measurement, giving only partial information.
\item Individual gates are ephemeral and invoked via control signals that can execute different gates and circuits.
\item Quantum noise, qubit decoherence, and typical faults are very different from any phenomena observed in CMOS circuits. In particular, stuck-at faults run against the unitary evolution implemented by quantum circuits.
\end{itemize}

\begin{figure}[tb]
\centering
\includegraphics[angle=0,width=0.8\linewidth]{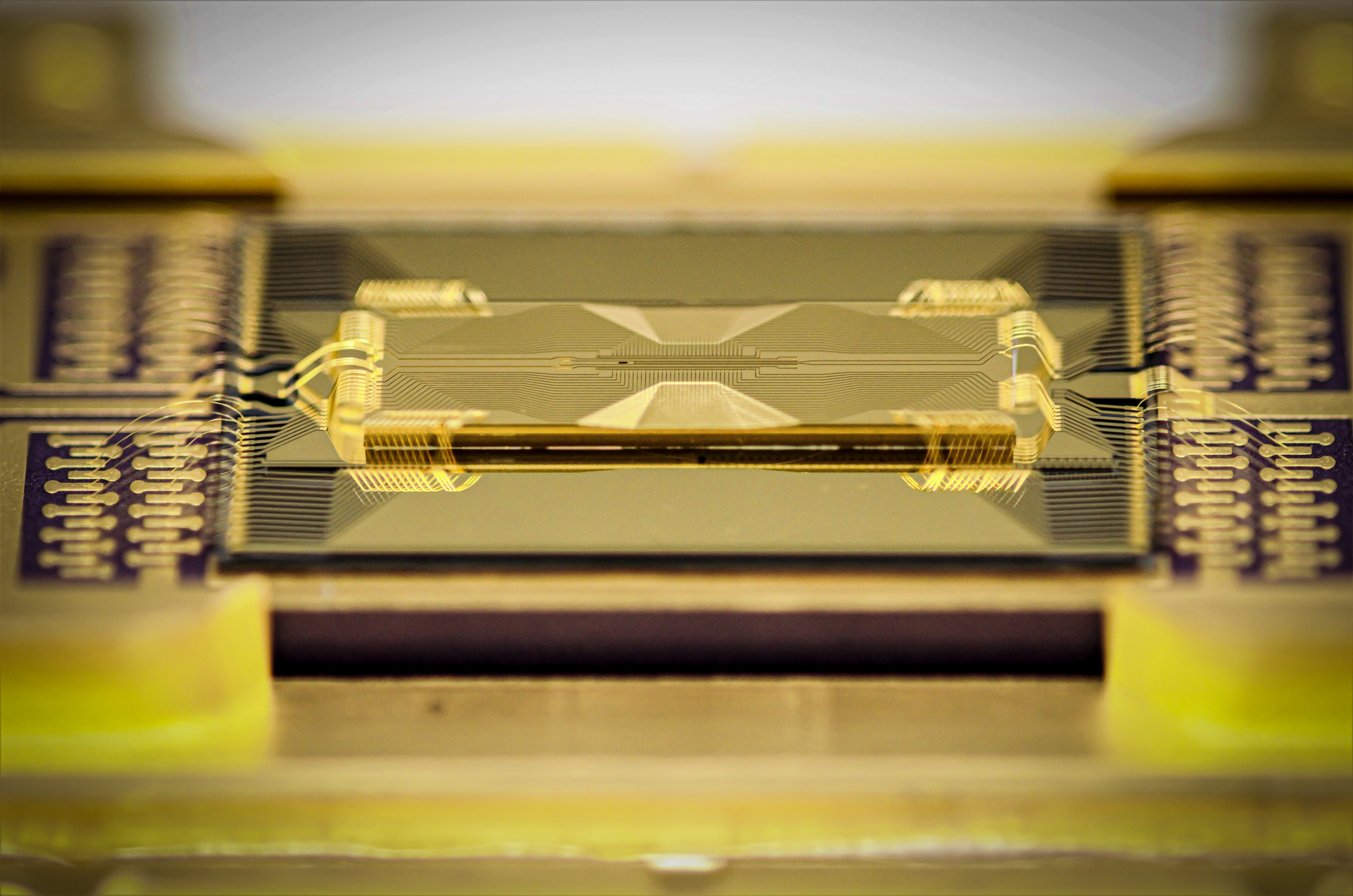}
\caption{\label{fig:trap_pic} 
A 14mm long, high optical access, \ionq-designed trap packaged on an interposer. Ions are loaded by ionizing a neutral beam of $Yb$ atoms introduced through a load slot in the left region of the trap. The ionized atoms are laser-cooled and then shuttled to the quantum zone in the center of the image. 
A qubit ion, excited by a laser, fluoresces (at UV wavelengths) or not depending on the quantum state, which helps perform quantum measurement.
Image credit: \ionq. 
}
\vspace{-3mm}
\end{figure}

Compared to scanning many test inputs into a static CMOS chip and observing intermediate values, quantum computers are tested by feeding an all-zeros initial state to a variety of test circuits which are reconfigured in an FPGA-like fashion. Observed measurement results are nondeterministic, which requires multiple repetition to collect sufficient statistics.

Whereas conventional computing is dominated by CMOS-based ICs, many promising quantum computing hardware platforms are currently in active R\&D. Quantum bits (qubits) can be carried by ($i$) photons, ($ii$) ions suspended in vacuum, and even ($iii$) quantized currents in superconducting loops. This work focuses on commercial\footnote{Here {\em commercial} means supporting paying customers~\cite{IonQ_Q3_2021}.} quantum architectures where trapped ions are suspended in a vacuum by electric fields~\cite{RevModPhys.62.531}. The ions are laser-cooled to form a crystalline structure, 
typically a linear chain in the so-called linear Paul trap~\cite{doi:10.1063/1.1658153} 
(Figure~\ref{fig:trap_pic}).
This architecture limits noise that may disturb quantum information, and leading implementations report a stability window (coherence time)
of $10^{11}$ sec for amplitudes ($T_1$ time) and $90$ min for phases ($T_2$ time)~\cite{Wang_2021}.\footnote{Long $T_1$ and $T_2$ coherence times are crucial for reliable quantum computation. Short coherence times limit the depth of quantum computations as information is lost to decoherence. Quantum error correction (QEC), if applied before a decoherence event occurs, can extend computation time, but today's physical systems lack sufficient resources to implement effective QEC.} 

Among the advantages of ion-trap quantum computers (QCs) are the perfect replicability of qubits within a device, all-pair qubit connectivity, as well as clean and relatively fast readout~\cite{IEEE}.
Unlike many competing architectures, where qubits need to be manufactured, ion qubits are naturally available, as atoms, and are identical every time.
Operations on atoms such as creating ions by stripping electrons from neutral atoms, processing quantum information encoded in the internal states of the ion qubits, and reading out the states, are performed with laser beams~\cite{Wright_2019}, a mature technology that has been around for decades.
Spectroscopic qubit readout is fast and clean and one can faithfully determine the measured state of the qubit~\cite{PhysRevLett.113.220501}.  Unlike with superconducting qubits, any pair of ion qubits can be directly coupled with no overhead to shuttle the quantum information around.  
 While ion-trap QCs have much slower gates, their advantage in qubit lifetimes and all-pairs connectivity compensates for this when running quantum algorithms, as shown in a detailed comparison of two leading quantum architectures \cite{Linke_2017}.

\begin{figure}
\centering
\includegraphics[width=1\linewidth]{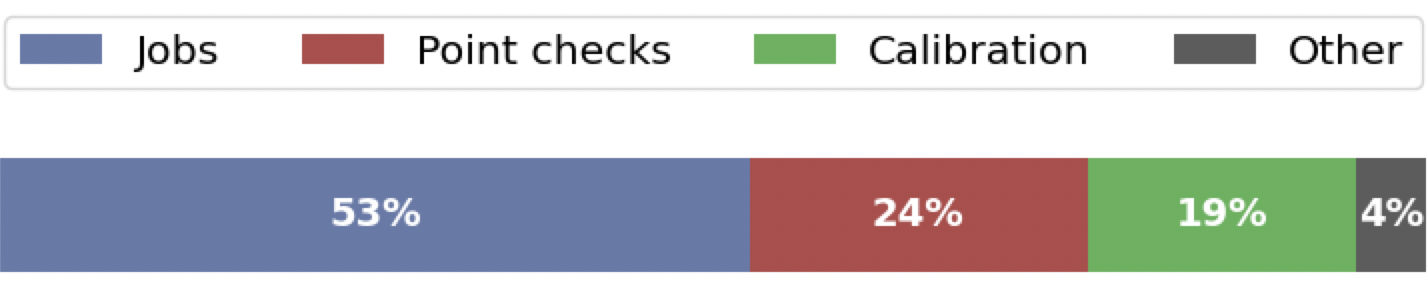}
\vspace{-4mm}
\caption{\label{fig:barchart} A typical duty cycle of a contemporary commercial ion-trap QC.
Roughly 53\% of time is spent to run QC jobs. The remaining 47\% is used to test and calibrate various device aspects, including qubit couplings.
}
\vspace{-3mm}
\end{figure}

 With the power to couple arbitrary pairs of qubits, comes the responsibility to ensure that qubit couplings are properly calibrated and can perform accurate gate operations~\cite{Maksymov_2021}. Therefore, the duty cycle\footnote{Here, {\em duty cycle} refers to the duration over which the ion chain can be used as a quantum computer. When the chain is lost, a new chain must be loaded and settled. This time is excluded from the duty cycle.} of an ion-trap QC interchanges operational periods and rounds of recalibration. Figure~\ref{fig:barchart} outlines the duty cycle of a commercial ion-trap QC. {Testing and calibration take almost half of the time for our QCs that run client jobs ($20\%$ for the next-generation, 21+ qubit machines~\cite{zhu2021generative,qedcArxiv} or $25\%$ for Honeywell architectures~\cite{Pino2021}), with significant overhead in identifying miscalibrated qubit couplings.} For each new check, different control signals must be uploaded and measurements must be repeated to collect statistics. In more mature QC systems with improved coupling quality, more subtle faults will complicate fault detection. {Calibration overhead correlates with  connectivity and makes up only $17\%$ of the duty cycle for the nearest-neighbor 2D array of superconducting qubits~\cite{isca2021,Arute2019}.} Testing $\binom{N}{2}$ qubit couplings, now already at a quarter of the ion-trap QC duty cycle,
 is going to consume a larger fraction of time as QC systems scale up. Hence, the need to optimize fault detection.
 
 In this paper, we aim to improve the ion-trap duty cycle and speed up recalibration by efficiently diagnosing faulty qubit couplings. Brute-force diagnosis that checks qubit couplings one at a time scales poorly and is typically limited to testing a small subset of couplings and recalibrating them as necessary. Testing multiple couplings at a time scales better, can identify faulty couplings, and offers two avenues for improvements: (1) faster recalibration of faulty couplings, (2) delayed calibration by mapping quantum circuits around detected faulty couplings.
 To this end, our key contributions are
 \begin{itemize}
     \item A classification of quantum faults in ion-trap QCs.
     \item Fault modelling for most common faults in the commercial architecture we explore.
     \item Testing protocols that quickly identify faulty qubit couplings among $\sim N^2$ possibilities using only $O(\log N)$ tests, always faster than binary-search strategies.
     \item A three-way validation of the protocols --- via proofs, on a commercial QC, and using simulation to show scaling.
 \end{itemize}
   
  In the remaining part of the paper, Section \ref{sec:background} reviews abstract quantum circuits and how they are implemented on ion-trap architectures. Section \ref{sec:faults} offers a classification of faults in ion-trap QCs and focuses on modelling the dominant fault types in near-term implementations. Section \ref{sec:problem} formalizes the problem we solve in Section \ref{sec:protocols}, where we propose fault-testing protocols for diagnosing the faulty qubit couplings. In addition to proofs of correctness, we then offer validation based on experiments with a commercial ion-trap QC (Section \ref{sec:experiment}) and simulation (Section \ref{sec:simulation}).
  
\section{Background}
\label{sec:background}

The promise of quantum computing is not in higher device density or more computational steps per second, but rather in a new model of computation that requires fewer computational steps to solve some tasks \cite{NC}. To this end, we review quantum circuits and outline how they are implemented on ion-trap QCs.

\subsection{Abstract quantum circuits}
\label{sec:qubits}

 Quantum information is represented in states
 of quantum systems, such as collections of interacting trapped ions~\cite{Wright_2019} or other physical systems. Instead of 0-1 bits,
 quantum bits (qubits) are used. In isolation, each qubit represent a two-dimensional complex-valued vector
 $(\alpha,\beta)$ with $|\alpha|^2 + |\beta|^2 =1$.
 The {\em computational basis} states $(1,0)$ and $(0,1)$ are shortened to $\ket{0}$ and $\ket{1}$, respectively, and a general one-qubit state is written as $\ket{\psi} = \alpha\ket{0} + \beta\ket{1}$~\cite{NC}.
 Measuring $\ket{\psi}$ w.r.t. the basis states changes it to either $\ket{0}$ or $\ket{1}$ with probabilities $|\alpha|^2$ and $|\beta|^2$, respectively. Physical details of ion-trap qubits are discussed in Section \ref{sec:traps}, but the formalism holds for many technologies, such as superconducting and photonic qubits.
 
 Quantum and conventional information differ in many ways, e.g., note a continuum of possible qubit configurations from which conventional
 information is extracted by destructive measurements. Another key distinction is in how subsystems can interact with each other. Conventional bit configurations are concatenated as strings, and the composite space grows as $2^n$ for $n$ bits. Quantum bits can be concatenated, e.g., $\ket{0}\ket{1}$ can be written as $\ket{01}$, but for superposition qubit we have:
 \[
 \big(\frac{\ket{0} + \ket{1}}{\sqrt{2}} \big)
 \otimes
 \big(\frac{\ket{0} + \ket{1}}{\sqrt{2}} \big)
= 
\frac{1}{2} (\ket{00} + \ket{01} + \ket{10} + \ket{11}).
 \]
 The resulting superposition is clearly {\em separable}, but removing one term, say, $\ket{01}$
 revokes separability and creates an {\em entangled state}. Quantum computing cannot outperform conventional computing without entanglement because separable computations can be simulated on conventional CPUs well.
 
 Quantum circuits typically start at a fixed state, such as $\ket{000\ldots0}$ (sometimes shortened to just $\ket{0}$) and apply a sequence of allowed operations, which end with qubit measurements~\cite{IEEE}. Before measurement, an ideal quantum computer preserves the state-vector norm and only applies {\em unitary} operators $U$.\footnote{Unitary operators are those satisfying $UU^*=I$. Equivalently, unitary operators preserve norms and dot products.} Most technologies today directly support only operators that modify one or two qubits at a time. Such operators are described by templates called {\em quantum gates} which can be instantiated on any qubit or pairs of qubits. Single-qubit gates include inverters (X gates) that swap $\ket{0}$ and $\ket{1}$, Z gates that negate $\ket{1}$, Phase gates that replace $\ket{1}$ with 
 $i\ket{1}$, and Hadamard gates that create superpositions of basis states
 $\sigma_x = X = \begin{bmatrix*}[r] 0 & 1 \\ 1 & 0 \end{bmatrix*}
 \ \ \ 
 \sigma_y = Y = \begin{bmatrix*}[r] 0 & -i \\ i & 0 \end{bmatrix*}
 \ \ \ 
 \sigma_z = 
 Z = \begin{bmatrix*}[r] 1 & 0 \\ 0 & -1 \end{bmatrix*}
 \ \ \
 P = \begin{bmatrix*}[r] 1 & 0 \\ 0 & i \end{bmatrix*}
 \ \ \    
 H = \frac{1}{\sqrt{2}}\begin{bmatrix*}[r] 1 & 1 \\ 1 & -1 \end{bmatrix*}.
 $
 Unlike these discrete gates, three Pauli-axis rotations $R_x(\theta)$, $R_y(\theta)$, and $R_z(\theta)$ perform angle-$\theta$ rotations. 
 \[
R_x(\theta) = \exp(-i\theta X/2) = \begin{bmatrix*}[r] \cos{\frac{\theta}{2}} & -i\sin{\frac{\theta}{2}} \\ -i\sin{\frac{\theta}{2}} & \cos{\frac{\theta}{2}} \end{bmatrix*}
\]
\[
R_y(\theta) = \exp(-i\theta Y/2) = \begin{bmatrix*}[r] \cos{\frac{\theta}{2}} & -\sin{\frac{\theta}{2}} \\ \sin{\frac{\theta}{2}} & \cos{\frac{\theta}{2}} \end{bmatrix*}
\]
\[
R_z(\theta) = \exp(-i\theta Z/2) = \begin{bmatrix*}[r] \exp(-i\frac{\theta}{2}) & 0 \\ 0 & \exp(i\frac{\theta}{2}) \end{bmatrix*}
\]
Any single-qubit unitary operator can be written as a product of three axial rotations, whereas the gates X, Y, Z, P can be seen as special cases of these rotations.
However, single-qubit gates alone cannot create entanglement. Common entangling two-qubit gates are Controlled-NOT (CNOT) and Controlled-Z (CZ)~\cite{NC}. They do not modify the {\em control} qubit, but apply either X or Z on the {\em target} bit subject to $\ket{1}$ on the control bit. For example, CNOT maps 
$\frac{\ket{0} + \ket{1}}{\sqrt{2}} \otimes \ket{1}$ to $\frac{\ket{01} + \ket{10}}{\sqrt{2}}$. Given that X = H Z H, we also note that $\mathrm{CNOT} = (I \otimes \mathrm{H}) \mathrm{CZ} (I \otimes \mathrm{H})$. Therefore, direct hardware support is often provided for only one of CNOT and CZ.

\subsection{Ion-trap quantum computers}
\label{sec:traps}

In a ${}^{171}{Yb}^{+}$ ion, computational basis states $\ket{0}$ and $\ket{1}$ are encoded in the energy levels $\ket{F=0, m_F=0}$ and $\ket{F=1, m_F=0}$ of the $^2S_{1/2}$ ground state, respectively.

\noindent
{\bf Initialization and readout} of typical ion qubits are performed using laser beams. Prior to any quantum computational operations, the ion chain is trapped and laser-cooled to the motional ground state using an efficient cooling method~\cite{PhysRevA.102.043110}.  As a result, a chain of room-temperature ions can be brought down to $\mu K$ temperatures in tens of milliseconds. {\it Optical pumping} then initializes a qubit to the $\ket{0}$ state with high accuracy in $\sim 20\mu$s. During measurement, laser light resonant with the $^2S_{1/2}$ $\ket{F=1}$ to $^2P_{1/2}$ transition induces fluorescence, allowing each ion to be imaged by individual photodetectors when illuminated, resolving to either the $\ket{1}$ or the $\ket{0}$ state, which typically takes $\sim 100\mu$s. Fast photo-detectors have been demonstrated with crosstalk between adjacent ions  $<10^{-4}$~\cite{Crain2019}.

\noindent
{\bf Quantum gates}, applied to evolve the quantum state, are implemented in an ion-trap quantum computer by illuminating the ions by another set of laser beams.
Modulating the beams helps implement single-qubit gates, where the illuminated ion's state is gradually rotated on the {\em Bloch sphere} about the desired Pauli axis. More complicated dynamics and modulation \cite{Zhu_2006,Leung_2018,PhysRevLett.114.120502,blumel2019poweroptimal} are used to realize the canonical M\o{}lmer-S\o{}rensen (MS) two-qubit gate \cite{PhysRevLett.82.1971,PhysRevA.62.022311}, which implements an ${\rm XX}(\theta) = \exp(-i\theta \sigma_x\otimes\sigma_x/2)$. This gate uses vibrational modes of the ion chain as the medium of information exchange, akin to a memory bus. 
A CNOT gate can be expressed via an MS gate and single-qubit gates as follows~\cite{Maslov_2017}: $\mathrm{CNOT} = (R_y(\pi/2) \otimes I) (R_x(-\pi/2) \otimes R_x(\pi/2)) {\rm XX}(\pi/2) (R_y(-\pi/2) \otimes I)$.
Clearly, the MS gate and single-qubit gates provide universality, i.e., can implement an arbitrary quantum operator~\cite{universal}. Typical gate {\em fidelity} ( Section~\ref{sec:faults}) of an MS gate is $\sim 96.5\%$ , whereas for a single-qubit gate it is $\sim 99.5\%$. The best reported fidelities are orders of magnitude better~\cite{PhysRevLett.113.220501,Gaebler2016, Ballance2016}, suggesting room for near-term improvement. 

\noindent
{\bf All-pairs qubit connectivity} is common for ion-trap architectures and allows any two qubits to be entangled directly. This is natural since each and every ion participates in a vibration bus. All-pairs qubit connectivity provides a decisive advantage in harnessing the computational power of an ion-trap QC~\cite{Grzesiak_2020},
and often the feasibility of a given computation on a given QC~\cite{Linke_2017}.
Applying fast gates to a remote qubit pair requires substantially greater power in the driving signals than for a nearby pair~\cite{blumel2021efficient}, which may increase the risk of crosstalk. Therefore, shorter connections are preferred.

\noindent
{\bf Control accuracy} required
to perform high-fidelity ion-trap quantum computation~\cite{Li_2020} is assessed in terms of the trapping field and the beams that illuminate the qubits. The former ensures proper geometric placement of trapped ions and with suitable vibrational behaviors. The latter ensure the alignment of optical signals with qubits.
Control loss occurs in several ways.
Stray fields in the ion trap are a common problem, one of leading causes is incidental charging of trap electrode surfaces. Even very small stray fields interfere with the trapping electric field and slowly displace ions from their ideal locations, indirectly degrading the quality of quantum gates. Understanding these mechanisms suggests restorative action. Since typical operation of an ion-trap quantum computer intentionally applies extra electric fields to the trap electrodes, we can use them to return ions to their optimal positions. This recalibration of ion positions can compensate for slowly-varying stray fields. 
Gate operations on ion qubits are performed using a pair of overlapping laser beams, requiring accurate control of beam geometry and power. Optomechanical stability is thus needed in order to avoid positional and power drifts. Additionally, the light used to illuminate the qubits results in significant light-induced static and dynamic frequency shifts of the qubit states. These {\em light shifts} are challenging to model~\cite{PhysRevA.94.042308}.

\ \\

\noindent
{\bf Component recalibration} is performed in two phases: measurement and correction. In one example, we gradually shift the beam position and monitor ion response.
In addition to geometric calibration, ion-qubit gates are also routinely calibrated to compensate for any optical power drifts. Respective light shifts are measured and accounted for in the execution of control algorithms. Unfortunately,
it is not practical to measure every relevant parameter, therefore recalibration is either repeated at regular intervals or triggered by a failing canary, i.e., a representative monitored value falling below an acceptable threshold.  
Lacking accurate up-to-the-minute information about physical drifts leads to significant recalibration efforts that are not entirely necessary.

\noindent
{\bf Benchmarking} a quantum computer assesses not only individual components but also their integration and the overall functional performance.
A full characterization of a quantum process in an $N$-qubit system requires exponentially large computational resources in $N$. Thus, system errors are evaluated by simplified methods, such as randomized benchmarking (RB) \cite{Emerson_2005}. RB essentially applies a random sequence of gates drawn from a restricted set of gates. Assuming the errors are non-systematic and Markovian, efficient estimation of the size of the errors associated with the gates in the set becomes possible. Multiple variations of RB are reported in the literature, such as cycle benchmarking (CB) \cite{Erhard_2019}, allow one to further and better characterize quantum gates and circuit elements with a more advanced and realistic model of errors.

\begin{table*}[tb]
\centering
\begin{tabular}{c|c|c}
         & \sc Unitary & \sc Non-unitary \\
    \hline
    \sc  Deterministic &  
      \parbox{7cm}{
      Inexact calibration of beam intensity, e,g, due to light shift miscalibration, beam misalignment or wrong gain applied to the illuminating beams. Usually static in time.}
    & 
    \parbox{7cm}{
     Non-unitary violations of physical models: unintended bit flips induced by signals that correspond to vibrational bus excitation, sidebands or anharmonicity. 
    }
    \\
    \hline
    \sc Stochastic    & 
    \parbox{7cm}{
    Random parameter fluctuations due to heating, control signal noise in amplitude and frequency.
    }
    & 
    \parbox{7cm}{
    Double ionization event, loss of order, chain loss, etc.
    }
    \\
    \end{tabular}
    \vspace{2mm}
\caption{
\label{tab:classification}
Types of quantum faults with respect to determinism and unitary evolution. A third axis accounts for the time scale.
}
\vspace{-3mm}
\end{table*}

\section{Fault characterization \& models}
\label{sec:faults}

 To address the non-ideal operation of a quantum computer, we classify such behaviors by their determinism and unitary evolution properties. An ideal QC should be deterministic before quantum measurement but this is not always the case. For example, qubit decoherence on QCs with low $T_1$ and/or $T_2$ times can limit the depth of quantum computation. The evolution of properly isolated quantum states, as described by Schr\"odinger equation, is unitary unless affected by stray interactions.
 Table \ref{tab:classification} discusses four types
 of quantum faults along these axes. A third axis in our classification accounts for the time scale. Slow noise may look deterministic during one QC run but not at a longer time scale. 

A QC free of noise and errors produces the correct output state which then determines the probabilities of measurement outcomes. For example, if a two-qubit circuit results in a fully entangled state $\frac{1}{\sqrt{2}}\left(\left|00\right>+\left|11\right>\right)$ then there should be a 50\% chance to measure the system in either $\left|00\right>$ or $\left|11\right>$ states while the probabilities to measure it in either $\left|01\right>$ or $\left|10\right>$ states should be zero. Errors present in real devices change the distribution of output probabilities~\cite{blumel2019poweroptimal,blumel2021efficient}, contributing to imperfect gate fidelities. 

\noindent
{\bf Gate fidelity estimation.}
Formally, gate fidelity ${\mathcal F}$ captures the similarity of an ideal unitary gate $U$ 
and the actual quantum operation
$\tilde{U}$ performed, i.e., ${\mathcal F} := 1- |U-\tilde{U}|$ for some operator norm $|\ldots|$. Popular choices include the Frobenius norm, as well as the diamond, spectral and other norms. In practice, 
estimating fidelity can be difficult because
the matrix $\tilde{U}$ is not easily available. Since the dominant failure modes of contemporary MS gates on ion-trap QCs are known, it is common to use approximate methods, which we illustrate next.

On an ion-trap QC, MS gates tend to have smaller gate fidelities than single-qubit gates. In the literature~\cite{blumel2019poweroptimal,blumel2021efficient}, MS gate fidelities are typically approximated by first applying $\mathrm{XX}(\pi/2)$ to $\left|00\right>$ and measuring ($i$) the populations of $\left|01\right>$ and $\left|10\right>$ and ($ii$) the balance between the populations of $\left|00\right>$ and $\left|11\right>$. Note that in the absence of inaccuracies the generated state should be
$\frac{1}{\sqrt{2}}\left(\left|00\right>-i\left|11\right>\right)$,
so that the lack of $\left|01\right>$ and $\left|10\right>$ and the balance between $\left|00\right>$ and $\left|11\right>$ are
indicative of perfect gate fidelity.

As a more detailed example, consider evaluating the MS-gate fidelity in the case where $\text{XX}(\pi/2)$ induces a perfect balance between $\left|00\right>$ and $\left|11\right>$. When the gate is applied to, say, ions $i$ and $j$, the average fidelity (over arbitrary two-qubit input states) may be computed as \cite{blumel2019poweroptimal,blumel2021efficient,PhysRevA.97.062325}
\begin{equation}
{\mathcal F} = 1 - \frac{4}{5}\sum_{p=1}^{n} [\eta_{p,i}^2 + \eta_{p,j}^2]|\alpha_p|^2,
\end{equation}
where $\eta_{p,i}$ is the so-called Lamb-Dicke parameter that denotes the coupling strength between the $p$th vibrational (normal) mode and the $i$th ion and $\alpha_p$ is the error incurred in the decoupling of the mode $p$ from any one of the ions at the end of an MS gate, required for a perfect MS-gate implementation, that corresponds to the amount of quantum information unintentionally left behind in a memory bus.\footnote{For completeness: $\alpha_p := \int_0^\tau g(t) e^{i\omega_p t} dt$, where $\tau$ is the MS gate duration, $g(t)$ is the control pulse illuminating the ions, and $\omega_p$ is the angular frequency of mode $p$. See \cite{blumel2019poweroptimal} for details.} In short, the {\em infidelity} $1-{\mathcal F}$ of an MS gate on an ion-trap QC is characterized by the residual, incorrect coupling between the ions $i$ and $j$ that the MS gate is applied to and the vibrational modes. The infidelity here can be directly measured as the sum of populations of $\left|01\right>$ and $\left|10\right>$.

Returning to a more realistic case with an imbalance between $\left|00\right>$ and $\left|11\right>$, the following expression is often used to compute the gate fidelity \cite{PhysRevLett.114.120502}:
\begin{equation}
{\mathcal F} = \frac{P^*_{\left|00\right>}+P^*_{\left|11\right>}+\Pi_{\rm contrast}}{2}.
\end{equation}
The two fidelity-determining circuits are $\mathrm{XX}(\pi/2)$ and $(R_\phi(\pi/2)\otimes R_\phi(\pi/2))\mathrm{XX}(\pi/2)$, where $R_\phi(\theta) = \exp(-i\theta (\cos(\phi)\mathrm{X} + \sin(\phi)\mathrm{Y})/2)$.
The initial state for both circuits is $\left|00\right>$.
Ideally, for the first circuit, as discussed above, we expect $P_{\left|00\right>}+P_{\left|11\right>} = 1$, where $P_{\left|ab\right>}$ are the populations of the states $\left|ab\right>$. We use $P^*$ to denote the measured populations from this first circuit.
For the second circuit, as a function of $\phi$, we ideally expect
$P_{\left|00\right>}+P_{\left|11\right>}-P_{\left|01\right>}-P_{\left|10\right>} = \sin(2\phi)$. If there is an imbalance between $\left|00\right>$ and $\left|11\right>$ that is induced due to implementing $XX(\pi/2+\epsilon)$, however, instead of $\sin(2\phi)$, one obtains $\cos(\epsilon)\sin(2\phi)$.
In practice $\Pi_{\rm contrast}\sin(2\phi)$ is observed, $\Pi_{\rm contrast} \leq \cos(\epsilon)$, since further contrast can be lost due to information leakage to the bus. Empirically, the value of $\Pi_{\rm contrast}$ is estimated via the best fit of ion-trap data.

The quality of these approximate methods can be tested directly by comparing theoretical predictions to experimental results. Results reported in the literature \cite{blumel2019poweroptimal,blumel2021efficient,PhysRevA.97.062325} show that the experimental results are in excellent agreement with the theoretical prediction. The agreement indicates that the failure modes of MS gates on today's ion-trap QCs are well understood, leading us to fault models we discuss next. 

\noindent
{\bf Dominant faults and fault models.}
Experimental data~\cite{Wright_2019} for leading ion-trap QCs suggest that, at this stage of development, unitary faults are most common and most impactful. Excellent isolation of trapped ions from non-unitary noise plays a big role here. In ion-trap QCs (and some superconductor QCs), coherence times $T_1$ and $T_2$ are orders of magnitude longer than the time needed to run a circuit~\cite{Wang_2021}, while state preparation and measurement errors (SPAM) are $<$1\% and can be addressed in post-processing due to their stability~\cite{ShenNJP2012,Ballance2016}.
On the other hand, the impact of unitary faults accumulates exponentially with the number of gates and modifies the final quantum state. 
When native gates are specified by their rotational axis and angle, most unitary errors impact the phase and/or the amplitude of a particular gate. Hence, our general fault models for single- and two-qubit unitary gate errors presented in Figure~\ref{fig:errors}. Other unitary errors such as spillover (cross-talk) can be simulated by additional gates applied on {\em spectator} qubits. The residual coupling of the chain to the motional modes can also be simulated by additional random single-qubit rotations. Fault models in Figure~\ref{fig:errors} explain how faults accumulate under gate repetition: deterministic faults add up until a certain point (based on how unitary rotation matrices multiply).

\begin{figure}[tb]
  \centering
  \includegraphics[width=0.7\linewidth]{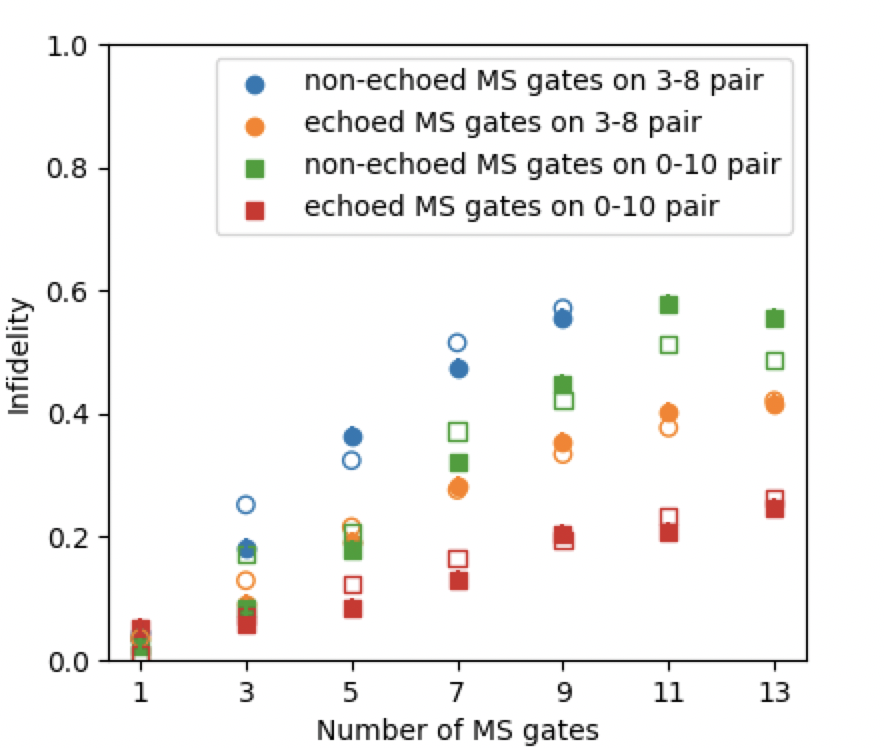}
  \caption{\label{fig:examp_conc_ms} Measured (solid markers) and simulated (empty markers) infidelities of concatenated MS gate sequences for $\{3,8\}$ (circles) and $\{0,10\}$ (squares) qubit pairs. Each circuit was measured in 1000 shots. In the non-echoed concatenated sequences, all MS gates have the same phases, while in the echoed sequences, gate phases shift by $\pi$ for every next gate.}
  \vspace{-3mm}
\end{figure}

\noindent
{\bf Empirical validation of fault models: simulation vs physical observations.}
Phase and amplitude errors can be non-deterministic (due to noise and drifts) and deterministic (due to miscalibrations and alignment errors). Due to the high dimensionality of the quantum state space, non-deterministic errors accumulate at a significant rate,
but slower than correlated deterministic errors~\cite{PhysRevA.92.042301}.
We illustrate this in Figure~\ref{fig:examp_conc_ms} by simulation and by a direct experiment on an ion trap. Shown are  infidelities of a sequence of MS gates concatenated in phase vs. in anti-phase for two ion pairs of an 11-ion chain. It can be noticed that gate fidelities differ between the two ion pairs due to the difference in noise and calibration errors. Our simulation models phase noise and residual coupling to the motional modes.
As seen in Figure~\ref{fig:examp_conc_ms},
our simulator and our fault models
show reasonable agreement with experimental data. 
 The dominance of deterministic unitary errors (systematic amplitude errors, etc.) gives us hope to find them with tests and eliminate them with recalibration. Multi-gate sequences can mitigate such errors~\cite{merrill2012progress,PhysRevA.90.040301,Grzesiak_2020,blumel2019poweroptimal}, but they often are unsuccessful given other error sources.

\begin{figure*}[!tb]
\resizebox{\textwidth}{!}{
$
R(\theta,\phi) = \begin{bmatrix}
\cos \frac{\theta}{2} & \hspace{-1em} -i e^{-i\phi} \sin \frac{\theta}{2} \\
-i e^{i\phi} \sin \frac{\theta}{2} & \cos \frac{\theta}{2}
\end{bmatrix}
$
\ \ 
$M(\theta,\phi_1,\phi_2) = \begin{bmatrix}
\cos \frac{\theta}{2} & 0 & 0 & \hspace{-1em} -i e^{-i(\phi_1+\phi_2)} \sin \frac{\theta}{2} \\
0 & \cos \frac{\theta}{2} & \hspace{-1em} -i e^{-i(\phi_1-\phi_2)} \sin \frac{\theta}{2} & 0 \\
0  & \hspace{-1em} -i e^{i(\phi_1-\phi_2)} \sin \frac{\theta}{2} & \cos \frac{\theta}{2} & 0 \\
-i e^{i(\phi_1+\phi_2)} \sin \frac{\theta}{2} & 0 & 0 & \cos \frac{\theta}{2}
\end{bmatrix}
$
}
\caption{
\label{fig:errors}
General one-qubit gates and two-qubit MS gates with small parameter values used as fault models for unitary errors in ion-trap QCs.}
\end{figure*}

\section{QC Testing: problem formulation}
\label{sec:problem}

Thanks to all-pairs qubit connectivity, $\binom{N}{2}$ different MS gates can be implemented on an $N$-qubit ion-trap QC. 
We consider the case where a small subset of qubit couplings are faulty and require calibration --- other couplings maintain their earlier calibration. Based on the characterized faults that dominate today's ion-trap QCs, we seek a testing strategy to find the faults quickly.

A typical ion-trap QC runs 1- and 2-qubit gates much faster than it can re-initialize qubits (using laser cooling, etc). Therefore, the runtime of a test with multiple MS gates is dominated by qubit initialization and readout, and we minimize the number of tests. For example, we can run a test circuit with one MS gate per coupling, excluding one particular qubit pair. If this test produces expected results, it would implicate the excluded qubit pair.
More efficient strategies use {\em binary search} where each test narrows down the set of suspected couplings by a factor of two.
Such strategies can succeed with arbitrary balanced partitions of couplings, requiring $\log_2 {\binom{N}{2}} \approx 2\log_2 N - 1$ tests to identify a single faulty coupling (the worst case is only one test worse than the average case). As seen in Section \ref{sec:faults} and Figure \ref{fig:errors}, quantum gates may be "slightly faulty", and very small inaccuracies can be neglected. 
To make sure that above-threshold faults are detected, we repeat each gate several times in each test.

Binary search is an {\em adaptive} testing strategy, where the results of prior tests determine the next test. However, such feedback and any required conventional computation to select the next test may be prohibitive without thorough hardware support because many {\em non-adaptive} tests can be performed before the next adaptive test is determined and loaded into the ion trap controls. For similar reasons, circuit ATPG for conventional VLSI is largely non-adaptive. Therefore, one of our challenges is to develop efficient non-adaptive strategies, even if they require more tests than binary search.

Depending on how often we run the tests, multiple qubit couplings may need calibration. Binary search finds one fault among many, but also extends to multiple faults. In one such approach, all couplings with detected faults are removed from future tests, and binary search is repeated to find the remaining faults.\footnote{A different strategy would perform two tests at each step --- testing a set of couplings and its compliment. If both tests detect faults, the search procedure would branch two-way. But this appears less efficient when the number of faults is small but unknown.} To avoid the expense of adaptations, we develop non-adaptive test strategies for single and multiple faults.

\section{Fault testing protocols}
\label{sec:protocols}

In this section, we show that for multiple faults adaptive testing can be done more efficiently than with binary search, and non-adaptive tests can do most of the job.
As is typical for textbook analysis of binary search, we simplify notation and analysis by first assuming that the number of qubits is a power of two $N = 2^n$. The more general analysis immediately follows by padding.
For brevity of analysis, we assume that when a coupling is not faulty, all gates on this coupling enjoy perfect fidelity.

\subsection{Prerequisite combinatorics} 
\label{sec:combinatorics}

We index individual qubits with integers $0..2^n-1$ and view these integers in binary. In this section, we investigate pairs of {\em distinct} integers $0..2^n-1$ and whether they belong to certain large classes of integers. 
We separately study 
\begin{enumerate}
\item {\em generic} pairs of integers with at least some bits in common, 
\item pairs of integers that share bits in specified positions (or none) and have opposite bits in the remaining positions.
\end{enumerate}

Define $2n$ classes labeled $(i,b)$, so that each class contains all integers whose $i$-th bit value is $b$, $i \in \{0,1,..,n-1\}$ and $b \in \{0,1\}$, i.e., $(i,b)$ is a $n{-}1$-subcube of an $n$-{\em Boolean cube} with $i$-th bit value $b$~\cite{knuth2014art}. 

\begin{lemma}
\label{lem:allpairs}
Consider pairs of distinct non-complementary integers. Each pair is included in at least one class.
\end{lemma}

\noindent
{\sc Proof sketch}: to find such a class for a pair, it suffices to find a common bit shared by the two integers.
\begin{lemma}
\label{lem:oneof}
For each $0\leq i < n$, the classes $(i,0)$ and $(i,1)$ are complementary, and every pair of integers is in at most one of them.
\end{lemma}

Since different $n$-bit integers can share at most $n-1$ bits, 

\begin{lemma}
\label{lem:pair}
Any two distinct integers are in
$\leq n-1$ classes.
\end{lemma}

\begin{example}
$ $ \\
$n=3$, the $(i,\cdot)$ classes are
\begin{centering}
\resizebox{0.4\columnwidth}{!}{
\begin{tabular}{c|c|c}
\hline
$i$ & $b=0$ & $b=1$\\
\hline
0   & 0, 2, 4, 6 & 1, 3, 5, 7\\
1   & 0, 1, 4, 5 & {\bf 2}, 3, 6, {\bf 7}\\
2   & 0, 1, 2, 3 & 4, 5, 6, 7\\
\hline
\end{tabular}
}
\end{centering}
\\

The integers $\{2,7\}$ share bits at $i=1$
and belong to the class $(1,1)$. But $\{0,7\}$, $\{1,6\}$, $\{2,5\}$, and $\{3,4\}$ are bit-complementary and do not belong to any class.
\end{example}

We now turn to pairs of bit-complementary integers and define a set of $2n-2$ classes. Class labels will be $[i,=]$ and $[i,\neq]$ for $0< i < n$. The classes contain integers with equal and unequal bits in positions $i-1$ and $i$, respectively.\footnote{
Observe that $[i,=] = (\mathrm{GrayCode}(i), 0)$
and $[i,\neq] = (\mathrm{GrayCode}(i), 1)$. See \cite{knuth2014art} for Gray code.} For such classes, we have 

\begin{lemma}
\label{lem:compl_pairs_classes}
For each $0<i<n$, each bit-complementary pair of integers is included in one of $[i,=]$ and $[i,\neq]$.
\end{lemma}

\noindent
{\sc Proof sketch}: For two bit-complementary integers and any pair of bit indices $0<i<n$, the $(i-1)$-th and $i$-th bits are in the same relation ($=$ or $\neq$) for both integers.

\begin{example}
$ $ \\
$n=3$, the $[i,\cdot]$ classes are
\begin{centering}
\resizebox{0.4\columnwidth}{!}{
\begin{tabular}{c|c|c}
\hline
$i$ & $=$ & $\neq$\\
\hline
1   & 0, 3, 4, 7 & 1, 2, 5, 6\\
2   & 0, 1, 6, 7 & 2, 3, 4, 5\\
\hline
\end{tabular}
}
\end{centering}
\\

The bit-complementary integers $\{2,5\}$ have unequal bits at $i=2$ and $i=1$ (01 and 10, respectively), thus belong to the class $[2,\neq]$.
\end{example}

\begin{theorem}
\label{thm:compl_pairs_unique}
No two bit-complementary pairs of integers belong to the same set of classes. The same holds for classes $[i,=]$ only.
\end{theorem}

\noindent
{\sc Proof sketch}:
Given two different bit-complementary pairs,
we can find an $0<i<n$ such that the bits at positions $i-1$ and $i$ of these pairs exhibit different ($=$ or $\neq$) relations. To make this argument constructive, we represent each pair by $n-1$ bits computed as $\oplus$ (exclusive OR) of consecutive bits (same result for both integers), and then find different bits between the two pairs. For example, $(2,5)$ would be represented by $11_b$. Since the classes $[i,=]$ and $[i,\neq]$ are complementary, their membership information is redundant.

For a subset of $k<n$ bit indices and bit values, we now turn our attention to integers that have specified values at the $k$ bit positions. We then consider pairs of such integers that are bit-complementary in the remaining bits. To adapt the construction of classes $[i,\cdot]$, we renumber the bits so that the $k$ bits go first. Hence, 
$n-k-1\leq n-1$ classes distinguish any two pairs,
per Theorem \ref{thm:compl_pairs_unique}.

\subsection{A single-fault protocol}
\label{sec:single}

We now show how to identify a single faulty qubit coupling with $3n-1$ tests and a single round of 
adaptation after $2n$ tests.
For each class, we perform one test that includes gates for each coupling between qubits in that class. Per Lemma \ref{lem:allpairs}, this tests all couplings except for those between qubits with binary-complementary indices. After performing these tests, we call the set of failing tests $(i,b)$ a {\em syndrome}.
Lemmata \ref{lem:allpairs}--\ref{lem:pair} 
yield the following:

\begin{corollary}
 Given a faulty coupling defined by a pair of distinct integers, a syndrome includes $\leq n-1$ failing tests, with no repeating $i$ values.
\end{corollary}

\begin{lemma}
\label{lem:manypairs}
For a syndrome of length $0<L<n$,
$2^{n-L-1}$ pairs of integers result in the same syndrome.
\end{lemma}

\noindent
{\sc Proof sketch}: Fixing $L$ bits out of $n$ leaves $2^{n-L}$ integers. For each such integer, we complement all $n-L$ bits not fixed to obtain a pair with exactly $L$ shared bits. Thus, there are $2^{n-L}/2$ pairs.

\begin{theorem}
\label{thm:3n1} A single faulty coupling can be found with $3n-1$ tests and one round of adaptation after $2n$ tests.
\end{theorem}

\noindent
{\sc Proof sketch}: determining the syndrome takes $2n$ non-adaptive tests. For a syndrome of length $L$, Lemma \ref{lem:manypairs} shows that $2^{n-L-1}$ pairs exhibit the same syndrome. To find one pair out of $2^{n-L-1}$, binary search needs $n-1$ additional adaptive tests. Instead, we use the $[i,=]$ classes defined earlier, adapted to $k=n-L$ bits not specified by the syndrome.

\begin{example}
For $n=3$, the observed syndrome in the first round
is $(0,0)=\{0,2,4,6\}$ and $(1,1)=\{2,3,6,7\}$ (hence, $L=2$). This specifies two bits $*10_b$ and leaves $2^{n-L-1}=1$ possibility: $\{2,6\}$; $*$ are don't-cares~\cite{knuth2014art}. However, in case the observed syndrome is only $(0,0)$ (hence,
$L-1$ and $**0_b$), we have two possibilities: $\{0,6\}$ and $\{2,4\}$. To tell them apart, we test for $\{0,6\}$ that have two equal leading bits.
\end{example}

\begin{corollary}
\label{cor:3n1} The claim of Theorem \ref{thm:3n1} holds even if some qubit couplings are not used.
\end{corollary}

\noindent
{\sc Proof sketch}: We exclude unused couplings from
the test sets defined in Sections \ref{sec:combinatorics} and \ref{sec:single}.
Faults on remaining couplings are still distinguished by the tests.

Now that we have defined the tests and proven that they can distinguish any two faulty couplings, we need an algorithm that finds the unique faulty coupling given test results. Fortunately, the number of couplings is small enough to evaluate test results for each and compare them to observations.

\subsection{Diagnosing multiple faults}
 With multiple faults, the first round of the single-fault protocol produces the union of syndromes of individual faults. For multiple pairs, the claims of Lemmata \ref{lem:oneof} and \ref{lem:pair} will often not hold and further steps of our single-fault protocol won't find even one pair. Here we note that test-driven calibration makes little sense after catastrophic effects with numerous faults. Hence, we assume few faults and diagnose them.

\begin{center}
\fbox{
\parbox{7.8cm}{
 The key principle is to {\em use lightweight tests to separate faults in time and magnitude before trying to diagnose them}, whereas diagnosed faults can be separated by qubit couplings.}
}
\end{center}

 Before discussing specific strategies, we remind the reader that the time overhead of running a circuit (that is not too deep) on an ion-trap QC is dominated by qubit initialization and readout rather than by gate count, whereas deciding which circuit to run based on the output of earlier circuits is much more costly than running a circuit.

\begin{figure} [!tb]
    \centering
    \tikzstyle{decision} = [diamond, draw,
    text width=5em, text badly centered, node distance=2.5cm, inner sep=0pt, fill=black!7]
\tikzstyle{block} = [rectangle, draw,
    text width=7em, text centered, rounded corners, minimum height=4em, fill=black!7]
\tikzstyle{line} = [draw, very thick, color=black!50, -{Stealth[length=4mm, width=3mm]}]
\tikzstyle{cloud} = [draw, ellipse, node distance=0.5cm,
    minimum height=2em, fill=black!7]
\scalebox{0.5}{    
\begin{tikzpicture}[, node distance = 0.5cm, auto]
    \large
    % Place nodes
    \node [block] (init) {canary test};
    \node [decision, below of=init, node distance=3cm] (faulty) {faulty?};
    \node [block, left of=faulty, node distance=5cm] (increase) {increase gate repetitions};
    \node [block, below of=faulty, node distance=3cm] (syndrome) {$2n$ tests};
    \node [decision, below of=syndrome, node distance=3cm] (check) {check syndrome $L$};
    \node [block, left of=check, node distance=5cm, label=above:adaptation] (adapt) {run $n-L-1$ extra tests};
    \node [block, right of=check, node distance=5cm] (reduce) {reduce gate repetitions};
    \node [block, below of=check, node distance=3cm] (recal) {calibrate faulty couplings from the relevant set$^*$};
    \node [cloud, right of=recal, node distance=5cm] (over) {start over};
    % Draw edges
    \path [line] (init) -- (faulty);
    \path [line] (faulty) -- node [, color=black] {no}(increase);
    \path [line] (faulty) -- node [, color=black] {yes}(syndrome);
    \path [line] (increase) |- (init);
    \path [line] (syndrome) -- (check);
    \path [line] (check) -- node [, color=black] {$L<n-1$} (adapt);
    \path [line] (check) -- node [, color=black] {$L>n-1$} (reduce);
    \path [line] (check) -- node [, color=black] {$L=n-1$} (recal);
    \path [line] (adapt) |- (recal);
    \path [line] (reduce) |- (init);
    \path [line] (recal) -- (over);
\end{tikzpicture}
}
    \caption{Starting with a canary test and aiming to detect the largest fault, we adjust the number of gate repetitions in the tests (e.g. 2MS, 4MS, etc). The number of gate repetitions needs to be increased if no faults get caught, and reduced if too many faults clutter the syndromes. The threshold is adjusted accordingly to maximize the fault vs no-fault contrast. After a faulty coupling is identified and recalibrated, it should be removed from the relevant set$^*$ for a period of time based on the typical calibration lifetime. This elimination allows for a sequential fault detection by the single-fault protocol.}
    \label{fig:test-flow}
\end{figure}
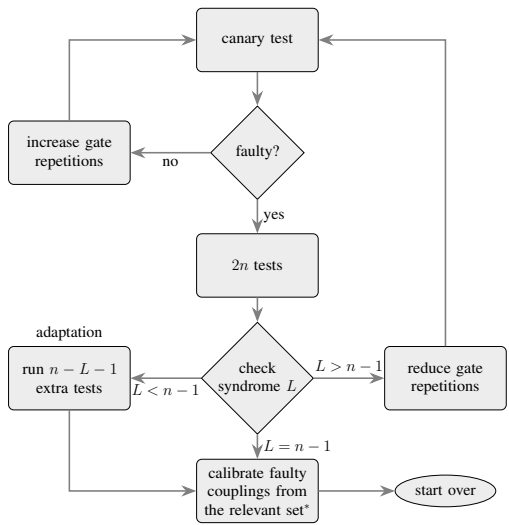

 \noindent
 {\bf Fault separation in time} relies on frequent (e.g., every minute) runs of a canary circuit with gates at all relevant qubit couplings, to detect the emergence of faults. 
 A canary failure triggers fault diagnosis (see Figure \ref{fig:test-flow}) before additional faults develop and complicate testing.

 \noindent
 {\bf Fault separation by magnitude.} As seen in Figure \ref{fig:errors}, faulty couplings may be off by different amounts. In practice, these amounts are often small initially but grow over time. To amplify small faults for detection, we repeat each gate several times\footnote{If the fault on a qubit coupling, upon repetition, becomes undetectable due to the fault's accidental cancellation or becoming identity, it may be detected by inserting qubit swaps. Example: suppose a coupling \{2,6\} is faulty such that applying the faulty operation twice in a row, applied to supposedly amplify the fault, cancels the effect of the fault. By inserting a swap between, for instance, qubits 5 and 6, the fault-detection test would then become calling (i) a faulty coupling between \{2,6\}, (ii) a qubit swap between \{5,6\}, and (iii) a non-faulty coupling between \{2,5\}, thus avoiding the accidental cancellation.}, and the number of repetitions is easy to control. An initial (canary) test is performed with every gate repeated many times. When it detects the presence of a fault, we find the smallest number of repetitions that still detects the fault. This can be done using a binary search on the number of repetitions or a non-adaptive search that checks $R$ different repetition counts but is likely to be faster in practice. When a single fault is associated with this magnitude threshold, it can be detected by the single-fault protocol.
 
 \noindent
 {\bf Fault separation by qubit couplings} removes the couplings with diagnosed faults from further testing and helps finding faults with smaller magnitude. Per Corollary \ref{cor:3n1}, the single-fault protocol works in this case, but
 we need to find the next smallest number of gate
 repetitions as explained above.
 
  \noindent
 {\bf The cost of fault detection} is dominated by adaptive rounds and is also (but less) sensitive
 to the number of qubit initializations and measurements, i.e., the number of circuit runs (including those with the same circuit). To this end, canary runs that provide {\em fault separation in time} can use delayed feedback, i.e., the next circuit run does not need to wait for canary feedback, but would be aborted in those rare cases when a fault is detected. {\em Fault separation in time} is performed by non-adaptive search (optimized using statistics of past fault magnitudes). A third round of adaptation is included in single-fault diagnosis. One additional round can verify that the identified fault is indeed faulty, after which we test for faults with a smaller magnitude. The overall procedure finishes when, after separating previously diagnosed faults, magnitude search finds no more faults (hence, another adaptation).
The costs are summarized as follows:

\begin{center}
\begin{tabular}{cc} 
\hline
  0 faults & periodic canary test runs (negligible) \\ 
  $k$ faults & $4k+1$ adaptations \\
\hline 
\end{tabular}
\end{center}

 The number of circuit runs is $ks(3n+R)$ where $s$ is the number of "shots" per circuit needed to collect statistics, and $R$ is the number of gate repetition configurations checked by non-adaptive search.

Combinatorial testing protocols have been widely considered in the literature for software unit testing~\cite{Nie2011}. However, our fault-testing protocols introduced in this work address the specific, dominant error models of trapped-ion QCs, differentiating them from techniques for conventional computing and even superconducting QCs. 

\section{Physical validation}
\label{sec:experiment}

In this section, we validate the fault-detection strategy detailed in the previous section. For this, we use an ion-trap QC hosted at \ionq. As a first step, we use a test that is readily available in experimental settings, called the {\em single-output test}. We investigate the test on well-pronounced artificial unitary errors that we introduce to the ion-trap QC and compare empirical results to those obtained from a unitary-error simulator, capable of simulating the kinds of errors that dominate today's ion-trap QCs. Upon confirming the agreement between the two, we again compare experimental data to and simulation results, however this time without introducing artificial errors. This directly targets the "natural" errors of ion-trap QCs. The observed agreement validates our error models, our fault-detection strategy, and its viability in practice.

\noindent
{\bf Ion-trap QC at \ionq.} We use an 11-qubit ion-trap QC for experiments. Each qubit is a ${}^{171}Yb^{+}$ ion, where the computational basis $\left|0\right>$ and $\left|1\right>$ is encoded in the hyperfine transition with splitting of $\sim 12.6$GHz. As explained in Section~\ref{sec:traps}, two-qubit gates are nominally calibrated to guarantee 96.5\% fidelity and single-qubit gate are nominally calibrated to guarantee 99.5\% fidelity. The qubits are roughly evenly spaced, with the inter-spacing of about $\sim 4\mu$m, resulting in $\sim 3$MHz frequencies of the motional modes we use as the medium of communication between qubits. $T_1$ time is immeasurably long (tens of thousands of years) and $T_2$ time is around 600ms. We note that while this system is limited in its scale, the working principles and the observed fault modes are representative of future ion-trap systems. Given the long $T_1$ and $T_2$ times, with the latter having been demonstrated to be more than an hour in the state-of-the-art experiment, we continue to expect unitary errors will be dominant for the foreseeable future. 

  \noindent
 {\bf Single-output tests.} A single test corresponds to a single quantum circuit with multiple two-qubit gates. Here we need a circuit sensitive to the errors of interest, so we use the single-output test that (ideally) returns the system to its initial state. The test passes if the resulting state matches the initial state. 
 Our tests that target deterministic unitary errors, such as miscalibrations. For example, if a fully-entangling MS gate is applied four times in a row onto the same qubit pair, it does not change the initial state in the absence of errors, while it applies $\text{XX}(4\epsilon)$ gate if there is a miscalibration that results in error $\epsilon$ per MS gate application. This is the rationale for a four-MS-gate, single-output test. The deviation of the output population from the all-zero state, with the all-zero initial state, is then indicative of the miscalibrated gates. A similar single-output test can use only two MS gates per pair with the difference being that the output is no longer the initial, all-zero state but an inverted, all-one state.

 \noindent
 {\bf Testing artificially introduced errors.} Here, we show experimental results for physical 8-qubit testing with sets of two and four stacked MS gates, and compare them with the simulations. Specifically, in an ion-trap QC, we introduce artificial 47\% and 22\% under-rotations on $\{0,4\}$ and $\{0,7\}$ qubit pairs, respectively, by adjusting amplitude of the beams illuminated for the qubit pairs. Each circuit was executed (and the results were measured) 300 times.
 
\begin{figure}[bt]
  \centering
  \includegraphics[width=1.0\linewidth]{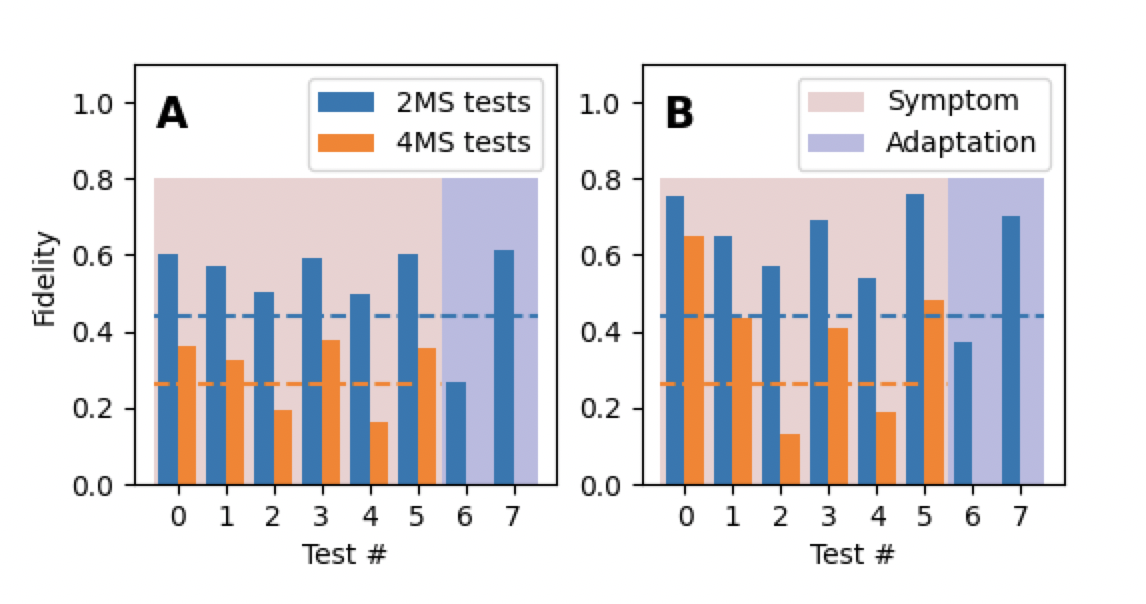}
  \vspace{-2em}
  \caption{\label{fig:ff_meas_xx} Measured fidelities of two-MS-gate tests (blue bins) and four-MS-gate tests (orange bins) with artificially-introduced 47\% and 22\% under-rotations on $\{0,4\}$ and $\{0,7\}$ qubit pairs, respectively, on 8 qubits. Panel A: Simulation. Panel B: Experiments. A positive test result can be determined by considering a fidelity threshold of 0.45 and 0.25 for the two- and four- MS-gate tests, respectively (dashed lines).}
\end{figure}

 Target-state fidelities, calculated from the returned measurement outcomes, are shown in Figure~\ref{fig:ff_meas_xx}. The obtained results show an excellent agreement with the simulated fidelities. Faulty tests can be resolved from non-faulty ones when compared to thresholds at $\sim0.45$ for the two-MS-gate tests and at $\sim0.25$ for the four-MS-gate tests. Since the main error source is of a unitary kind, we use a unitary simulator. In particular, in the simulator, we include 10\% random amplitude errors for all two-qubit gates, residual coupling to the motional modes that generates 1\% odd population (see Sec.~\ref{sec:traps}), and $1/f$ phase noise (see Fig.~\ref{fig:errors}). 

\begin{figure}[bt]
  \centering
  \includegraphics[width=1.0\linewidth]{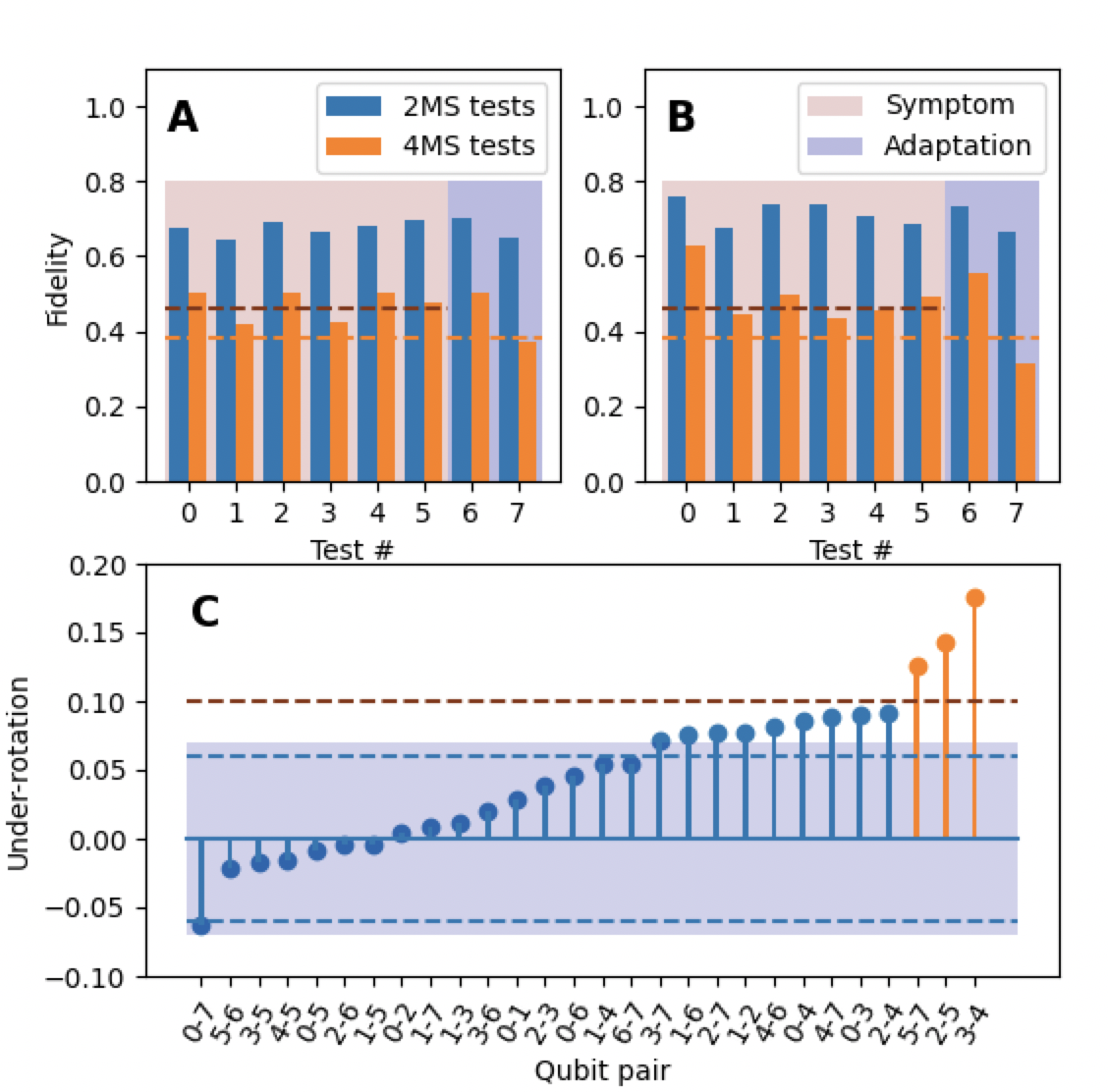}
  \vspace{-1em}
  \caption{\label{fig:ff_meas_xx_nonartificial} Measured fidelities of two-MS-gate tests (blue bins) and four-MS-gate tests (orange bins) with naturally occurred miscalibrations while idling for 15 mins. Panel A: Simulation. Panel B: Experiments. Positive test results for under-rotations of $\sim$15\% are determined by considering fidelity thresholds of 0.38 and 0.46 on four-MS-gate tests. Panel C shows the snapshot of the MS-gate quality at the 15th min. The blue area marks the region of under/over- rotations within 6\%.
  The red dashed line marks
  the 10\% under-rotaion threshold.}
\end{figure}

  \noindent
 {\bf Testing a commercial ion-trap QC in operation.}
Figure~\ref{fig:ff_meas_xx_nonartificial} shows a counterpart to Figure~\ref{fig:ff_meas_xx}, except, this time, we do not introduce artificial miscalibrations. Instead, we calibrate all our two-qubit couplings, wait 15 minutes, then run our tests. This serves as a litmus test on the efficacy of our fault-detection strategy under a typical operation condition of a commercial ion-trap QC. Independently, we directly monitor the MS-gate quality, i.e., we track angles $\theta$ of $\text{XX}(\theta)$ gates, for cross comparisons. The largest miscalibration on $\{3,4\}$ is diagnosed first with no positive test results.\footnote{This case results in no syndrome for the first six tests. The two tests we run after adaptation check for $01_b$, i.e., we see if there is any fault between $\{0,7\},\{1,6\}$ then $\{1,6\},\{2,5\}$. Thus, the two additional tests also return nothing. Assuming a single-fault model, the faulty coupling $\{3,4\}$ is identified. An additional test explicitly checking for $\{3,4\}$ can be run to rule out the zero-fault case.} Subsequently, the other two faults on $\{2,5\}$ and $\{5,7\}$ are diagnosed by inserting fidelity thresholds at 0.38 (dashed orange line) and 0.46 (dashed red line), respectively.

\section{Validation via simulation}
\label{sec:simulation}

\begin{figure}[bt]
  \centering
  \includegraphics[width=1.0\linewidth]{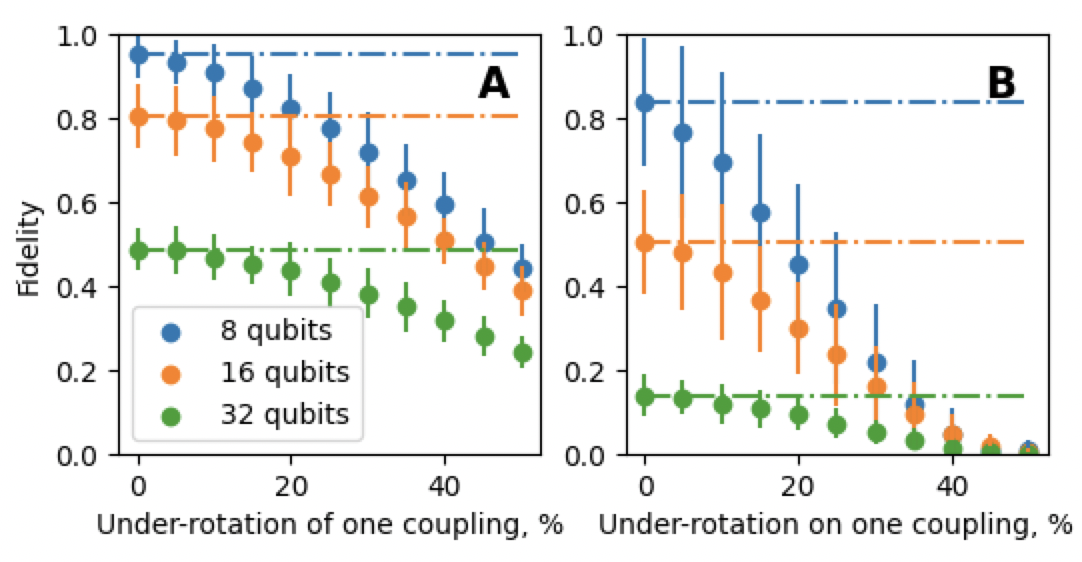}
  \caption{\label{fig:sim_scaling_xx} Simulated fidelities of two-MS-gate tests (panel A) and four-MS-gate tests (panel B) as a function of under-rotation on one qubit pair in the presence of 10\% average calibration error for 8, 16 and 32 qubits. Dashed lines extend the average fidelity absent calibration outliers.}
\end{figure}

\begin{figure}[bt]
  \centering
  \includegraphics[width=1.0\linewidth]{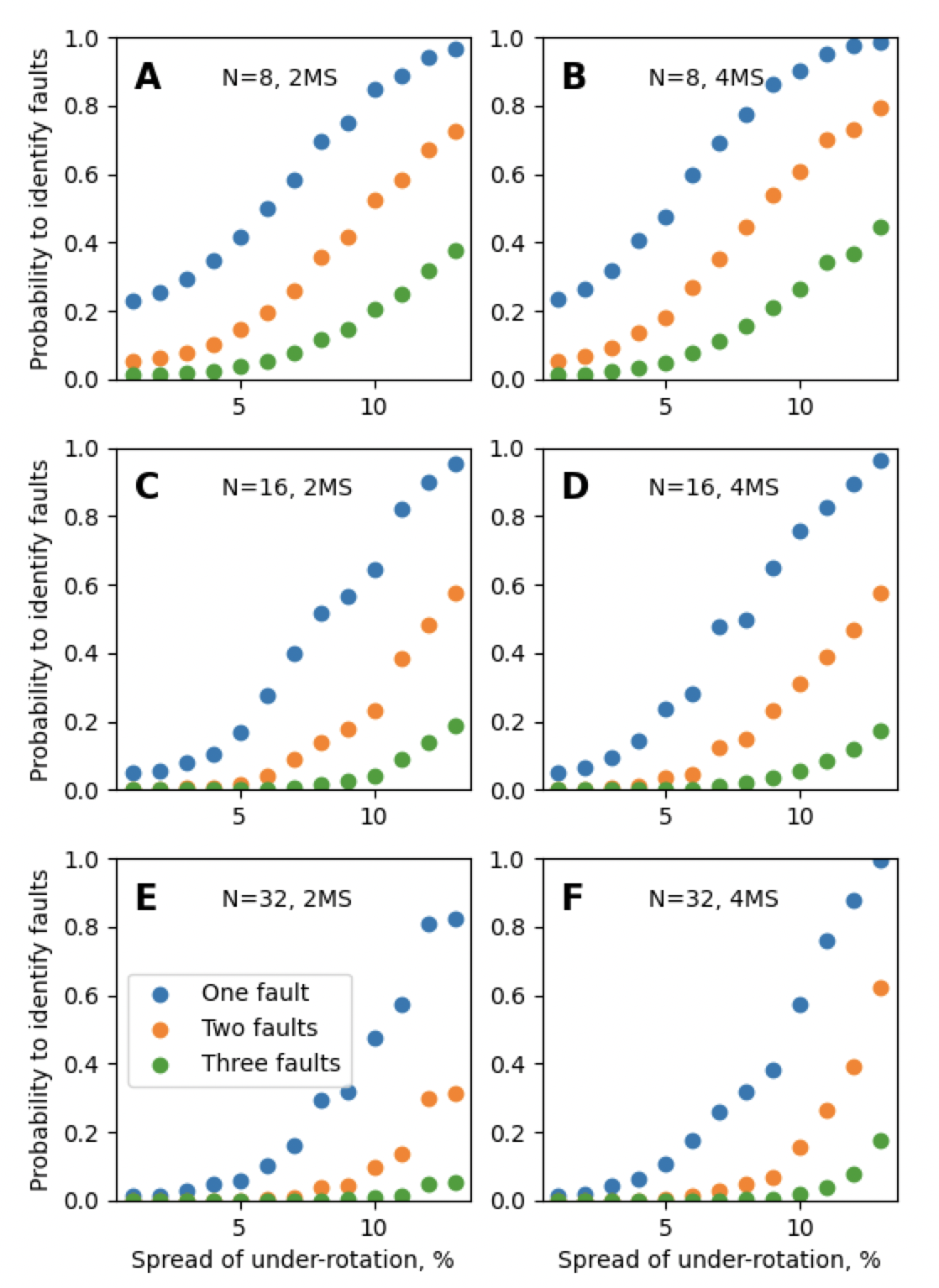}
  \includegraphics[width=1.0\linewidth]{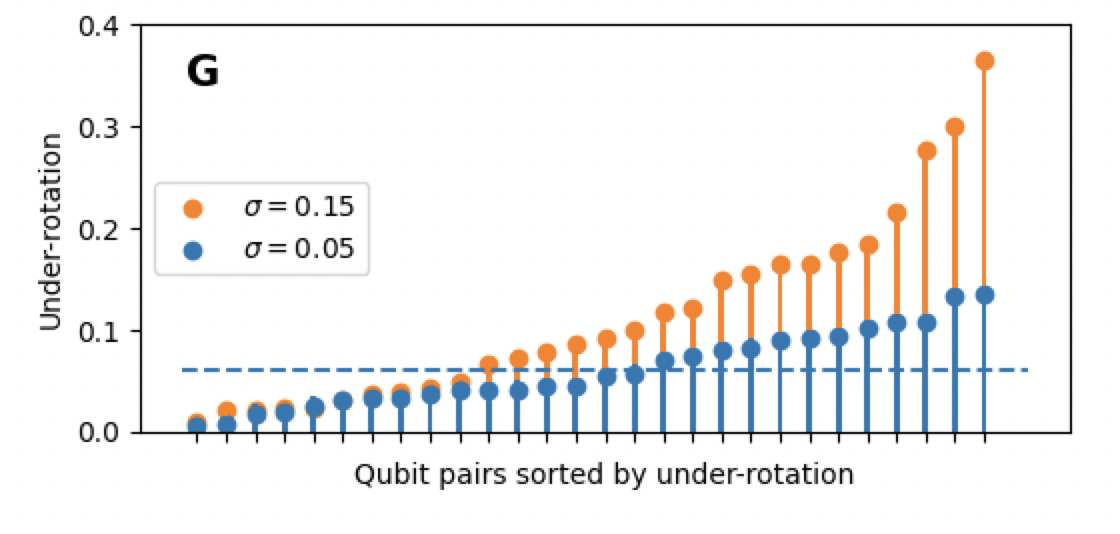}
  \vspace{-1em}
  \caption{\label{fig:ff_ideal_xx} Simulated probabilities to identify one, two and three faulty gates (panels A-F) depending on the Gaussian spread of under-rotated gates in the presence of uniformly spread under-rotation up to 6\% (dashed line on panel G) for 8, 16 and 32 qubits tested with two- and four-MS-gate tests. Two sorted generated distributions of gates by under-rotations are shown for $\sigma=0.05$ and $\sigma=0.15$ on panel G.}
\end{figure}

To go beyond the scale limitations of the physical QC available to us, we now leverage simulation to investigate the scaling of our fault-detection strategy. Here we use the same simulator that we confirmed (in Section \ref{sec:experiment}) to adequately model errors on our ion-trap QC (see Figure~\ref{fig:ff_meas_xx_nonartificial}).
Working with a simulator is particularly convenient, as we can create and explore in detail a variety of coupling fault configurations that (a) stress our fault-testing protocols and/or (b) represent realistic conditions in an ion-trap QC. For clarity of narrative, we suppress phase noise and residual couplings to the motional modes that we know to not affect the test outcomes in a significant way, leaving only 10\% random amplitude errors. We simulate up to 32 qubits with this error model.

Figure~\ref{fig:sim_scaling_xx} shows the fault contrast for larger numbers of qubits (8, 16, and 32), obtained from simulations of the two- and four-MS-gate tests. In the presence of 10\% amplitude noise, the faulty qubit pair, in the single fault case, needs to be an outlier to be distinguished through tests. For the two-MS-gate tests, the minimum under-rotations for 8, 16 and 32 qubits are about 25\%, 30\%, and 35\% respectively to be identified in 95\% of the cases. Under the same conditions, for the four-MS-gate tests, these numbers are lower by about 5\% (20\%, 25\% and 30\%), since deeper circuits show higher contrast. On the downside, average fidelities of the four-MS-gate tests decay faster with the number of qubits given other noise sources.

Figure~\ref{fig:sim_scaling_xx} shows that two faults that are too close in magnitude cannot be separated, which can scramble test syndromes.
Table~\ref{tab:faults} gives estimates of the probability to correctly identify faulty gates for 8, 16, and 32 qubits, based on how syndromes start repeating with the increased number of faults.

\begin{table}[t]
  \caption{Probabilities to identify faulty gates in the presence of one, two, and three faults on $N=$8, 16, and 32 qubits, where $N=2^n$.}
  \label{tab:faults}
  \centering
  \begin{tabular}{ccccl}
    \toprule
    n & Qubits &1 fault&2 faults&3 faults\\
    \midrule
    3 & 8 & 100\% & 47\% & 22\%\\
    4 & 16 & 100\% & 23\% & 5\%\\
    5 & 32 & 100\% & 12\% & 1\% \\
    \bottomrule
  \end{tabular}
\end{table}

Using the results of Figure~\ref{fig:sim_scaling_xx} and Table~\ref{tab:faults}, we estimate success probabilities of identifying faulty couplings for under-rotaion distributions similar to the experimentally observed distribution (Figure~\ref{fig:ff_meas_xx_nonartificial}C). 
Focusing now on the spread of under-rotations identified in the previous section, Figure~\ref{fig:ff_ideal_xx} shows the success probability to correctly identify the faulty coupling as a function of the \% spread of the under-rotation. Considered are both two- and four-MS-gate tests, with larger number of qubits (8, 16, and 32). 

We assume the following
distribution of faults by under-rotations. For $\leq$6\% under-rotations, we use a uniformly distribution. The choice of 6\% is motivated by the calibration threshold that can be used in practice. For larger under-rotations, we use a right-tail Gaussian distribution centered at 6\%, capturing the observed phenomenon of largely miscalibrated gates (orange points on Figure~\ref{fig:ff_meas_xx_nonartificial}C). The under-rotations sampled from this composite distribution are shown on Figure~\ref{fig:ff_ideal_xx} for $\sigma=0.05$ and $\sigma=0.15$.\footnote{Visually speaking, the composite distribution is flat up to 6\% at $a$ and then drops off according to the Gaussian distribution with the peak value of $a$. Normalization of the distribution function to one determines $a(\sigma) = 1/(0.06+\sigma\sqrt{\pi/2})$.} On average, as the spread increases, the errors are more separated by magnitude, the efficiency of detection increases not just for one but for multiple largest ones. The four-MS-gate tests show faster improvement due to higher contrast. However, the improvement is slightly impacted by the greater effect of the 10\% random amplitude noise on the four-MS-gate tests over their two-MS counterparts.

\section{Discussion}
As ion-trap QCs scale up to attain quantum advantage, their all-pairs qubit couplings give
a significant boost in computational power ~\cite{Maslov_2017}. However, those qubit couplings require individual calibration and tend to gradually lose calibration. Their recalibration
is already consuming a significant portion of the ion-trap QC duty cycle and will only grow as a fraction due to their $\binom{N}{2}$ scaling.

Our fault-detection strategy avoids costly recalibration of all couplings and often postpones recalibration since quantum circuits
can be mapped around several miscalibrated qubit couplings. Unlike binary search, our combinatorial testing approach almost entirely avoids adaptive tests.

Consider how an adaptation-based test  is performed:

\noindent
{\bf Step 1.}
A predetermined set of test circuits, each requiring qubit initialization, quantum gate execution, and qubit readouts, is run on a QC, repeated $r$ times.

\noindent
{\bf Step 2.}
A conventional computer receives the readout results, determines the error syndrome that may involve measurement-error correction on the readouts, and identifies the next batch of tests to be run on an ion-trap QC. 

\noindent
{\bf Step 3.}
An ion-trap QC controller receives the next test instruction, compiles the control pulse according to the instruction, and uploads the compiled pulse. 

\begin{figure}[bt]
  \centering
  \includegraphics[width=0.9\linewidth]{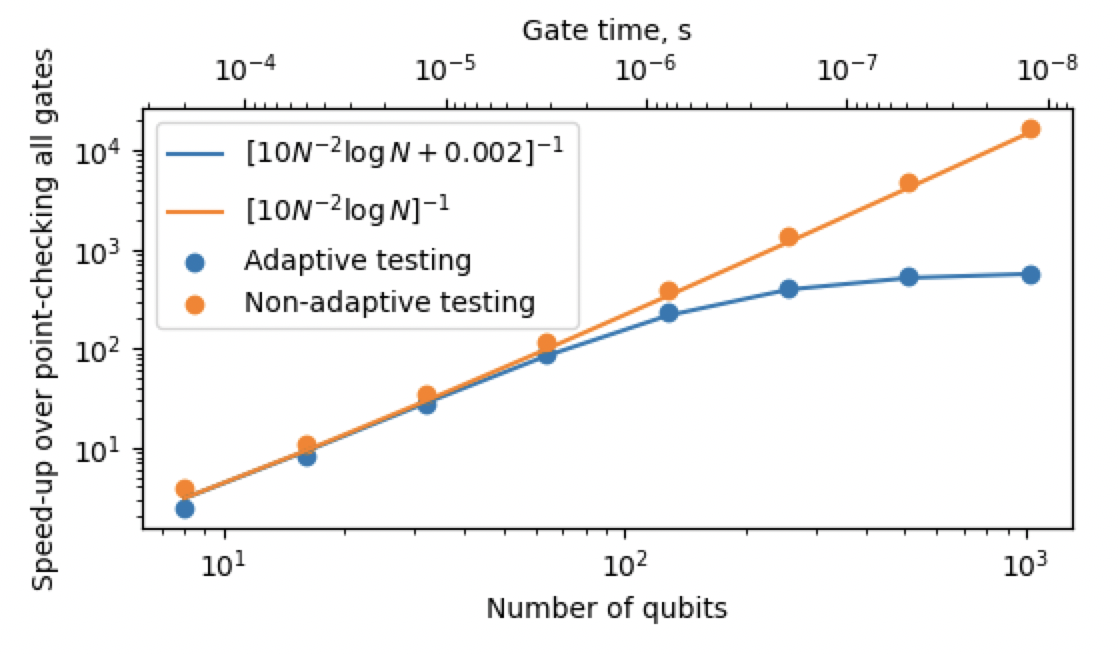}
  \vspace{0.5mm}
  \caption{\label{fig:comp_adapt} Speed-up for adaptive and non-adaptive testing compared to all-gate point-checks as a function of the number of qubits, $N$, assuming the gate time scales as $1/N^2$ starting with 0.2 ms for eight qubits. The speed-up for the adaptive testing plateaus due to the compilation time that depends on the number of couplings (blue line). It is still about $10^3$ times smaller than the processing time required per point-check. The speed-up for the non-adaptive testing is projected to grow as $N^2/\log N$ (orange line).}
    \vspace{1mm}
\end{figure}

The cost of adaptation can be estimated as the time taken by Steps 2 and 3, which may be significant compared to that of Step 1 (see Figure~\ref{fig:comp_adapt}), especially in the regime where quantum operations become faster, an expected future for ion-trap QCs.

\begin{figure}[bt]
  \centering
  \includegraphics[width=0.9\linewidth]{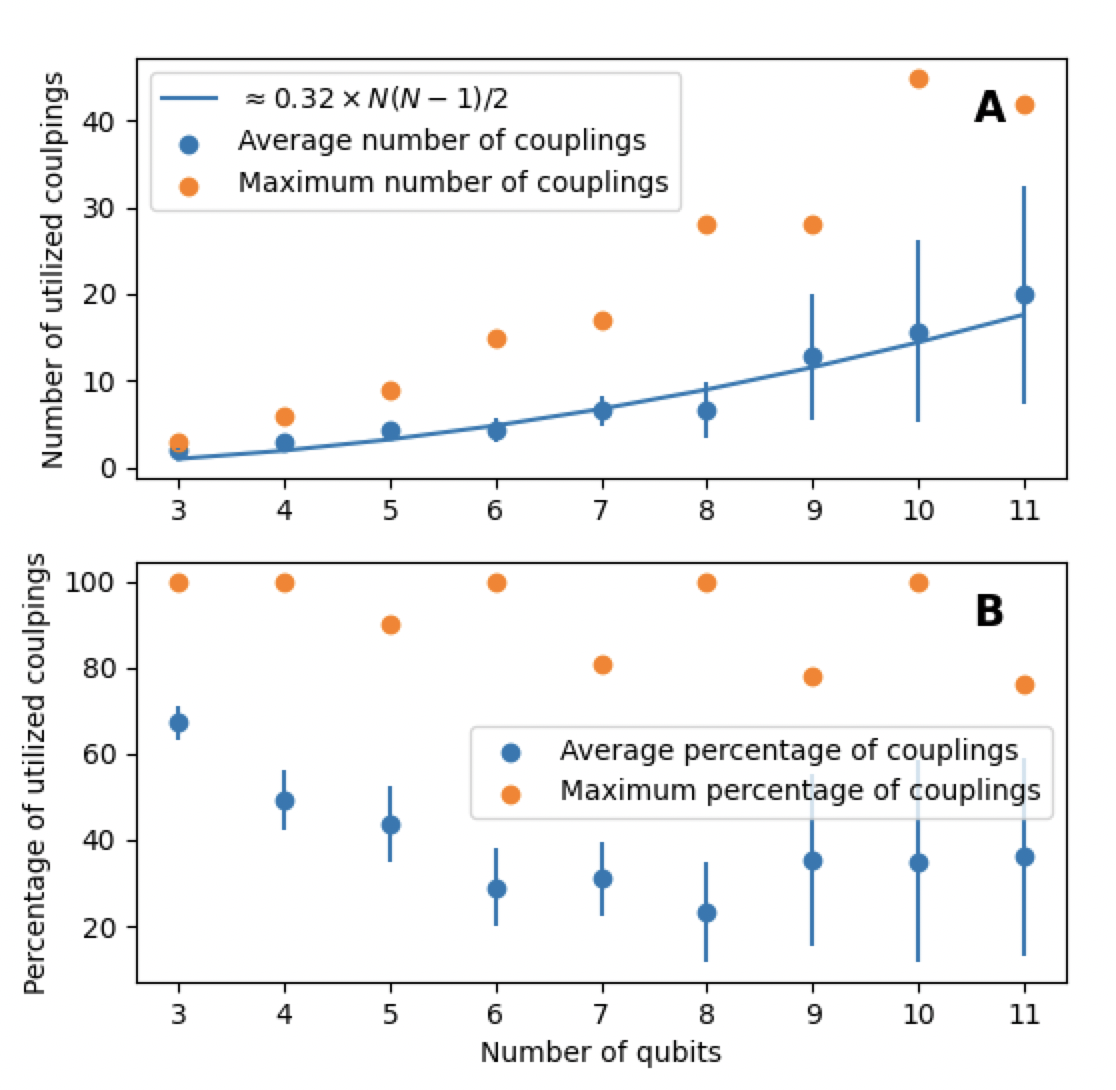}
  \vspace{0.5mm}
  \caption{\label{fig:gate_frac} Number of utilized couplings (panel A) and their fraction of the total number of couplings (panel B) as a function of the number of qubits used in real-life quantum circuits \cite{Maksymov_2021}. The average number of utilized gates scales as $\sim1/3$ of the total number of available gates (blue line).}
 \vspace{1mm}
\end{figure}

Some quantum circuits do not use all available qubit couplings (Figure~\ref{fig:gate_frac}). In this case, an identified faulty qubit coupling can simply be avoided from being used, instead of executing the calibration process for the coupling. In fact, more than one faulty qubit couplings can be tolerated, as more become detected, as per Corollary~\ref{cor:3n1}, until an input quantum circuit to run can no longer be adequately mapped to an ion-trap QC to avoid all known faulty qubit couplings. A related work that aims to constructively determine the minimal set of qubit couplings that needs to be calibrated for a given batch of input quantum circuits is available in \cite{Maksymov_2021}. Our detection-based strategy offers a complementary solution to reducing the resource overhead for qubit coupling calibrations. 

\section{Conclusions and outlook}

Ion-trap quantum computers are a fledgling technology with a multiyear track record of commercial cloud-based availability
and significant upside potential. Key advantages include (1) chip-based technology with no need for exorbitant cooling equipment, (2) long qubit decoherence times $T_1$ and $T_2$, (3) high-fidelity gates, and (4) the availability of all-pair qubit couplings. Near-term priorities include technology development, more accurate quantum control, and more effective duty cycle management. Our work improves the efficiency of periodic recalibration of qubit couplings by making them selective, based on an efficient fault-testing strategy. To enable this strategy, we developed a fault model for dominant faults. Our contributions are validated with mathematical proofs, physical experiments with a commercial ion-trap QCs, and numerical simulations that show a good agreement with physical experiments and attractive scaling with the number of qubits.

The main impact of our work is in increasing the uptime of commercial QCs by reducing the fraction of their duty cycle dedicated to recalibration. Compared to today's strategy that would take over a minute for a full characterization of every coupling, our strategy, applied to an 11-qubit system~\cite{Wright_2019} can diagnose the full system in ten seconds. While the number of qubits today is too small to fully illustrate the advantages of our diagnostic strategy, the advantage will grow more pronounced as ion-trap QCs scale. Indeed, a next-generation IonQ QC with over 21 qubits has recently been benchmarked by an independent third party~\cite{qedcArxiv}. The demonstrated scalability of commercial ion-trap QCs requires our efficient diagnosis strategy to ensure their viability, viewed through the lens of their duty cycle. Future revisions of ion-trap platforms will improve basic technology (lasers, detectors) and quantum control (electric fields, beam alignment and power), as well as ion-trap capacity. Under such technology scaling, it will be particularly important to maintain all-pairs qubit connectivity --- a major advantage of ion traps over solid-state quantum computing technologies that are limited to nearest-neighbor qubit couplings\cite{Linke_2017}.
As the number of qubits grows, the number of qubit pairs grows faster, making old-style recalibration unsustainable. Our test-based recalibration facilitates more attractive scaling.

\bibliographystyle{IEEEtranS}
\bibliography{citations}

\end{document}